# COMB-BASED RADIO-FREQUENCY PHOTONIC FILTERS WITH RAPID TUNABILITY AND HIGH SELECTIVITY

V.R. Supradeepa[1,2]†, Christopher M. Long[1,3]†, Rui Wu[1]†, Fahmida Ferdous[1], Ehsan Hamidi[1,4], Daniel E. Leaird[1] and Andrew M. Weiner[1]*

[1] School of Electrical and Computer Engineering, Purdue University, West Lafayette, Indiana, USA.

[2] Now at OFS Laboratories, Somerset, New Jersey, USA.

[3] Now at EPFL, Lausanne, Switzerland.

[4] Now at Harvard University, Cambridge, Massachusetts, USA.

†These authors contributed equally to this work.

*Corresponding author: amw@ecn.purdue.edu

Abstract

Photonic technologies have received considerable attention for enhancement of radio-frequency (RF) electrical systems, including high-frequency analog signal transmission, control of phased arrays, analog-to-digital conversion, and signal processing. Although the potential of radio-frequency photonics for implementation of tunable electrical filters over broad RF bandwidths has been much discussed, realization of programmable filters with highly selective filter lineshapes and rapid reconfigurability has faced significant challenges. A new approach for RF photonic filters based on frequency combs offers a potential route to simultaneous high **stopband attenuation, fast tunability, and bandwidth reconfiguration. In one configuration tuning of the RF passband frequency is demonstrated with unprecedented (~40 ns) speed by controlling the optical delay between combs. In a second, fixed filter configuration, cascaded four-wave mixing simultaneously broadens and smoothes comb spectra, resulting in Gaussian RF filter lineshapes exhibiting extremely high (>60 dB) main lobe to sidelobe suppression ratio and (>70 dB) stopband attenuation.**

Optical frequency combs, generated via self-referenced and stabilized mode-locked lasers, have enabled revolutionary progress in precision optical frequency synthesis and metrology[1-4]. Optical combs are also of tremendous interest for other applications[5], such as multi-wavelength coherent lightwave communications[6-8], optical arbitrary waveform generation[9-11], generation of low-phase-noise[12] or agile ultrabroadband microwaves[13], and signal processing[6,14]. For these purposes, in which higher pulse repetition rates are desired and only moderate frequency stability is required, comb sources based on strong electro-optic modulation of a continuous-wave laser have seen substantial attention[15-16]. Here we report significant advances in RF photonic filters enabled by the ability to rapidly tune the timing of the comb and shape its power spectrum.

In general, radio-frequency photonics has received considerable attention due to the low loss properties of optical fibers, which can be harnessed for transmission of RF signals (in optical form) over distances and bandwidths that would not be possible with RF cables, immunity to electromagnetic interference, and reduced weight[17-19]. For signal processing applications, the high carrier frequency is a fundamental consideration empowering photonics approaches – e.g., even a 20 GHz RF bandwidth constitutes only 0.01% of the optical center frequency. Although rapidly reconfigurable RF filters could have significant implications for concepts such as cognitive radio, in which the radio may tune adaptively to utilize free regions of the RF spectrum[20,21], tuning of RF filters over a large fractional bandwidth has traditionally been difficult. However, bandwidths that are large and challenging to work with in the RF domain are relatively small when translated into the optical regime and may be much easier to handle[22,23]. The introduction of MEMS variable capacitor approaches has led to recent advances in tunable RF filters[24-26]. The fundamental tuning speed of RF MEMS devices is characterized in [24] as limited to approximately 1-300 μs. Filters based on solid-state diodes have fundamentally higher tuning speeds but suffer from increased nonlinearity, loss, and power consumption[24]. In either technology challenges in controlling the coupled response of multiple resonances to maintain high filter selectivity practically constrain tuning to speeds much slower than the fundamental limits. In photonics one approach seeks to achieve RF filtering by transmitting RF modulation sidebands of an optical carrier through an optical filter. In one example, apparently based on a cascade of five coupled optical microresonators, a fifth-order filter with 70 dB stopband suppression[27] was reported. However, similar to filters implemented directly in the RF domain, tuning is complex because changes to the individual resonators must be synchronized to maintain the desired multipole filter function[28]. Rapid tuning of the RF filter response by tuning the laser frequency relative to the optical filter has been proposed[28]. However, rapidly (e.g., submicrosecond) controllable tuning of a CW laser over a range of a few GHz is itself a difficult challenge, and to the best of our knowledge, the use of this technique to realize rapid RF filter tuning has not been demonstrated.

A different and more heavily studied approach for RF photonic filters follows a tapped delay line architecture[17-19]. In contrast to coupled resonator approaches, tapped delay lines yield finite impulse response (FIR) filters characterized by the zeroes of the response function[29]. Such FIR filters are commonly considered for digital filter design and provide substantial flexibility for engineering both amplitude and phase response[30,31]. Broadband light sources such as mode-locked lasers[32,33] or continuous-wave incoherent light[34] have recently been exploited, in conjunction with dispersive propagation, to scale RF photonic tapped delay line filter implementations to a large number of RF taps with a single physical light source and a single delay element. Recently, the use of a recirculating frequency shifter as a tapped delay line yielded filters with passbands centered in the 100-200 MHz range, free spectral range (FSR) of ~35 MHz, main-to-sidelobe suppression ratio (MSSR) of 52 dB, and stopband attenuation of >70 dB[35]. However, tuning was not demonstrated. Finally, in [14] we demonstrated an approach for photonic implementation of RF electrical filters based on electro-optically generated frequency combs apodized using a programmable pulse shaper[36]. These filters have center frequencies and FSRs in the GHz range and bandwidths of hundreds of MHz, appropriate for channelization of RF spectra at resolution matched to electronic analog-to-digital converter technology. However, the full potential of optical frequency combs has not been explored.

Here we introduce several new concepts in comb-based RF photonics that bring special advantages for filter shape and tunability. First, careful tailoring of the comb spectral profile may be accomplished through the nonlinear interaction of a pair of synchronized, frequency-offset combs, directly as part of the generation process and without the need for a pulse shaper. The extreme smoothness of the resulting spectrum leads to tapped delay line filters with greatly enhanced MSSR. Second, the relative delay between a synchronized pair of combs can be switched in tens of nanoseconds, which translates into similarly rapid switching of the RF passband. An important point in our approach is that filter lineshape is in principle independent of filter tuning, which substantially eases control complexity. The possibility of bandwidth reconfigurability is also demonstrated. Comb-based RF photonic filter configurations offer the potential to provide simultaneously a high number of taps, fast tunability, bandwidth reconfiguration, and high stopband attenuation, a combination difficult to achieve with other known approaches.

**Results**

Figure 1(a) shows a schematic of an optical frequency comb based RF photonic filter. A comb of optical frequencies spaced by $\Delta f$ ~10 GHz is generated by periodic intensity and phase modulation of a continuous wave (CW) laser[15-16], then passed through a single sideband modulator driven by the RF signal that we wish to filter. This places an identical sideband onto each of the optical comb lines. Then the signal propagates through a dispersive fiber and is converted back to the RF domain using a photodetector. The comb lines arrive at the photodetector with different delays and hence function as independent taps in a tapped delay line filter architecture. The RF impulse response $h(t)$ and frequency response $H(\omega)$ can be written as[14]:

$$h(t) \propto \sum_{n=0}^{N-1} |e_n|^2 \delta(t-nT) \quad \text{and} \quad H(\omega) \propto \sum_{n=0}^{N-1} |e_n|^2 e^{-jn\omega T} \quad (1)$$

where $e_n$ is the electric field amplitude of the $n^{th}$ line of the comb, and $T=2\pi\Delta f\psi_2$ is the differential delay between comb lines, determined by the group delay dispersion $\psi_2$ and the frequency spacing between comb lines $\Delta f$. From (1) the RF filter has a periodic frequency response with an FSR of $T^{-1}$ and passband shape governed by the Fourier transform of the optical comb spectrum. By properly apodizing (or shaping) the comb spectrum, filters with high stopband attenuation and user defined passband functions may be achieved. Here we focus on filters with passbands that are approximately Gaussian, a shape that provides fast roll-off both in frequency and time domains. Previously we used a high resolution pulse shaper[9,36] to shape the comb spectrum, resulting in stopband attenuation down to ~35 dB, limited by 0.37 dB ripple in the apodized spectrum (please refer to Supplemental Information)[14]. Here we realize substantially improved passband shapes by introducing new methods for shaping of the comb directly in the generation process.

Figure 1(b) shows our setup for direct generation of an approximately Gaussian shaped comb spectrum, and Fig. 1(c) shows the principle. Three cascaded lithium niobate intensity modulators generate a train of 10-GHz pulses that are approximately Gaussian in time, to which we then apply a strong sinusoidal phase modulation. For pulsed signals centered at extrema of the sinusoid, the temporal phase is approximately quadratic, which results in time-to-frequency mapping, with the temporal intensity of the waveform mapped onto its power spectrum[37-38]. This time-to-frequency mapping enables direct generation of an approximately Gaussian frequency comb. Figure 1(d) shows the generated spectrum in this scheme; and Fig. 1(e) shows the frequency response of the RF photonic filter, measured using a vector network analyzer (VNA), and overlaid with a simulation based on Eq. (1) and the measured comb spectrum. As expected, we see two Gaussian-like passbands separated by the filter FSR. Fiber dispersion produces a tap delay of $T$ = 96 ps and a corresponding FSR of 10.4 GHz. The passband centered at 10.4 GHz has a 3-dB bandwidth of 750 MHz, with stopband attenuation of over 28 dB. Overall the measured filter function is close to the simulated response based on the comb spectrum, with confirms the predicted Fourier transform relationship.

**Rapidly frequency tunable RF photonic filters**

Although the passband frequency may be tuned in the scheme of Fig. 1 by changing the amount of dispersion, this is inconvenient and not rapidly adjustable. Recently we proposed a new approach to achieve filter tuning with fixed dispersion[14], shown schematically in Fig. 2(a). This method utilizes an interferometric configuration in which one sample of the comb is single sideband modulated with the carrier suppressed, while a second sample of the comb remains unmodulated but experiences a tunable optical delay. The frequency response of the RF filter obtained after the two samples of the comb are combined, sent down a dispersive line, and detected can be written as[14]:

$$H(\omega) \propto \sum_{n=0}^{N-1} e_{1n} e_{2n}^{*} e^{-jnT(\omega - \tau/\psi_2)} \qquad (2)$$

where $e_{1n}$ and $e_{2n}$ are the complex electric field amplitudes of the $n^{th}$ comb line in the two arms of the interferometer and $\tau$ is the relative delay between the two optical paths. If $e_{1n}$ and $e_{2n}$ are from the same comb, their phases will cancel, again leading to tap amplitudes given by $|e_n|^2$, i.e., by the comb spectrum. Equation (2) shows that the center frequency of the filter can be controlled by the relative delay. This was demonstrated for the first time in [14], which used a mechanical delay line to achieve quasi-static tuning.

In this paper we introduce a new scheme that exploits dual electro-optically generated frequency combs to achieve radically enhanced tuning speed, as shown in Fig. 2(b). A single CW laser is split and directed into two nominally identical comb generators, each following the approach of Fig. 1(b). As above, one of the combs is single-sideband modulated (but now with suppressed carrier), while the other remains unmodulated but is provisioned with controllable delay. A key observation is that since the two comb generators are driven at the same RF frequency, the relative delay of the generated optical trains is governed by the relative phase shift of the respective RF drive signals. Analog RF phase shifters are available that can vary phase over a $2\pi$ range with >50 MHz modulation bandwidth. By utilizing such a phase shifter, the delay of the generated comb can be electronically tuned over a full comb period in tens of nanoseconds. Figure 2(c) demonstrates rapid tuning of the optical delay between combs. The experiment is set up so that initially the two comb signals are overlapped in time (top panel). A step input to the control port of the RF phase shifter induces a change in the relative delay. As seen from the bottom panel, one of the optical pulse trains is shifted in this example by 20 ps (corresponding to ~20% of the comb repetition period) in < 30ns. This permits tuning of the RF filter response within a comparable time span, substantially faster than conventional RF tunable filter technologies with potential for high selectivity.

Figure 2(d) shows quasi-static tuning of the filter. By changing the control voltage into the phase shifter by only 0.75 V, the filter passband is shifted 3 GHz. The varying gain is attributed to the periodic optical filter used for carrier suppression in the SSB-SC block that also attenuates RF sideband signals that are close to the carrier. Gain variation could be substantially reduced by using a periodic optical notch filter with sharper edges, such as an etalon in reflection mode. Important points are (a) the passband shape, bandwidth, and stopband attenuation (~30 dB) are similar to the filter response of Fig. 1, and (b) these quantities remain largely unchanged when the filter is tuned. In accord with the predictions of eq. (2), the filter frequency may be tuned independently of the passband shape; this greatly simplifies high-speed tuning of such filters. This property has been demonstrated previously using microwave delay-line filters with complex coefficients[39,40], but implementation complexity limited the number of taps, and fast tuning was not reported.

The novel blend of electronic technologies and optics in our setup enables tuning simultaneously at high speed and over broad bandwidth. To view the response of such rapidly time varying filters, a time domain measurement technique (as opposed to pervasive frequency domain methods based on network analyzers) is absolutely required. An RF test waveform in which a 3.7 GHz tone and a 1.8 GHz carrier subject to a 600 Mb/s binary phase shift keying modulation are superimposed and input into the SSB-SC modulator, and a high speed control signal is connected to the control port of the phase shifter. Figure 3(a) shows a high level view of the RF signal at the output of the photodiode and of the control signal to the RF phase shifter (shifted to account for the filter latency). The control signal comprises three different voltage levels that tune the filter to three different center frequencies. Zoomed-in sections of oscilloscope traces of the filter output, corresponding to high and mid level control voltages for which 3.7 and 1.8 GHz input signals are individually selected, are shown in Figs. 3(b) and 3(c), respectively. The low control voltage tunes the filter to a frequency band not present on the input signal, and the filter output goes low. Figure 3(d) shows a spectrogram of the data set, which provides an intuitive visual representation of the time-frequency character of the filtered signal. The different frequency components superimposed at the input are now separated in time, and the presence of modulation on the 1.8 GHz signal is clearly evident as an increased bandwidth. Examination of the rapid transitions between output frequency components reveals that the filter is able to switch within 40 ns, limited by the response time of the RF phase shifter. Such rapid filter tuning is unprecedented in RF photonics and is substantially beyond the demonstrated capabilities of any known high selectivity filter technology.

**Bandwidth reconfigurability**

In addition to *tuning*, in which the filter passband is shifted rigidly in frequency, *reconfiguration* of the passband shape is also of interest. Here we demonstrate reconfiguration of the filter bandwidth. According to eq. (1), the RF bandwidth varies inversely with the bandwidth of the optical comb, which in turn is proportional to the RF drive signal to the phase modulator. By placing a variable attenuator before the phase modulator, as shown in Fig. 4(a), the generated optical bandwidth may be varied. Figure 4(b) shows measured RF transfer functions in a fixed filter configuration. By reducing the drive voltage to the phase modulator by approximately a factor of two, the optical bandwidth is decreased by a similar factor, resulting in an increase in RF bandwidth from 170 MHz to 360 MHz. The filter shapes remain similar, in each case with >30 dB MSSR. These proof-of-concept results employ a manually variable attenuator, allowing quasi-static reconfiguration; however, by upgrading to electronically controllable RF attenuators, bandwidth reconfiguration within tens of nanoseconds is anticipated. We also note that because this bandwidth reconfiguration scheme is fundamentally independent of the frequency tunability demonstrated earlier, in the future it should be possible to achieve simultaneous, independent, and rapid control of both passband frequency and bandwidth. This could be done by provisioning both of the "shaped combs" of Fig. 2(b) with electronically controllable RF attenuators so that their optical bandwidths can be modulated in identical fashion.

**Improving passband shape**

We now focus on a final important attribute for filter applications, namely selectivity, manifested through high MSSR and stopband attenuation. The key to high selectivity in our scheme is to achieve an extremely smooth and specifically shaped comb spectrum with minimal line-to-line variations. Direct electro-optic generation of an approximately Gaussian comb, described above, is a first step, but substantial improvement in spectral shaping is still required. Here we introduce a dual-laser driven comb generator plus nonlinear fiber optics approach that simultaneously broadens and smoothes the comb spectrum to achieve selective filters with greatly enhanced MSSR and stopband attenuation. High efficiency conversion through four-wave mixing has been previously utilized for a variety of tasks in optical communications[41-42]. Here we utilize cascaded four-wave mixing as a phase modulation amplifier and spectral smoother.

Figure 5(a) shows the experimental setup for the comb generator. As in Fig. 1(b), EO intensity and phase modulators produce a quasi-Gaussian spectrum on a CW laser centered at 1542 nm. A second CW laser (1532 nm) is combined without intensity modulation and passes through the same phase modulator. For reasons that will become clear, one comb is delayed with respect to the other by one half the modulation period (50 ps at 10 GHz). This is followed by an amplifier and zero dispersion, low dispersion slope highly nonlinear fiber (HNLF). Cascaded four wave mixing (FWM) in the HNLF leads to amplification of the phase modulation and substantial spectral broadening and smoothing. Figure 5(b) shows a schematic of the comb generation process. The initial two combs have narrow bandwidth and poor spectral quality. However, for successive FWM terms the bandwidth increases while the spectral shape becomes more Gaussian.

To understand this process more quantitatively, let $a_1=|a_1(t)|\exp(j\phi(t))$ be the complex amplitude of laser 1 (centered at $f_1$) which undergoes intensity and phase modulation. The intensity envelope $|a_1(t)|^2$ is shaped to look roughly Gaussian, and the temporal phase $\phi(t)$ is sinusoidal. Laser 2 (centered at $f_2$) is only phase modulated, and its complex amplitude can be written as $a_2=\exp(j\phi(t))$. After transmitting through a length of single-mode fiber (SMF) which delays laser 2 by half a modulation period with respect to laser 1, the complex amplitude of laser 2 can be written as $a_2=\exp(-j\phi(t))$. In the HNLF these waveforms mix to give a cascade of FWM terms. In the limit of negligible dispersion, the first FWM term centered at $(2f_1-f_2)$ will be primarily composed of

$$a_1^2 a_2^* = |a_1(t)|^2 \exp(j3\phi(t)) \tag{3}$$

Similarly, the 2$^{nd}$ order FWM will go as

$$a_1^3 a_2^{2*} = |a_1(t)|^3 \exp(j5\phi(t)) \tag{4}$$

Equations (3) and (4) correspond to scaling of the bandwidth by three times and five times, respectively. Furthermore, since the temporal intensity profile becomes increasingly localized as it is raised to progressively higher powers, the temporal phase is increasingly well approximated as quadratic. This results in higher fidelity time-to-frequency mapping and increasingly smooth combs.

Figures 6(a) and 6(b) show the comb spectrum of laser 1 right after the modulators and of the 2$^{nd}$ order FWM term, respectively. Both bandwidth enhancement and spectral smoothing are clearly seen. Figure 6(c) shows the corresponding filter transfer functions. The measured filter functions have different FSRs due to the 20 nm red shift of the 2$^{nd}$ order FWM signal and the dispersion slope of the DCF module. The filter response (blue dashed line) for the comb shown in Fig. 6(a) has a peak at baseband with 3-dB bandwidth of 350 MHz, while at 10.4 GHz there is a passband with 800 MHz 3-dB bandwidth. The filter response (red solid line) for the comb shown in Fig. 6(b) has a peak at baseband with 3-dB bandwidth of 210 MHz, while at 9.38 GHz (corresponding to 107.5-ps tap delay) there is a passband with 440 MHz 3-dB bandwidth. The roughly two-fold reduction in passband width is attributed to the increased comb bandwidth. Most importantly, the MSSR and stopband suppression are increased to 61 dB and >70 dB, respectively, the latter limited by the VNA noise floor. Simulations (see Supplementary Information) suggest that to reach these levels of performance, random variations of the comb spectrum must have standard deviation below ~0.2% ($9\times10^{-3}$ dB). Such variations are ~40 times smaller compared to previous results in which apodization of a comb is performed using a pulse shaper[14], leading to improvements in MSSR and stopband attenuation by >25 dB. The level of sidelobe suppression and stopband attenuation demonstrated here is extremely rare in RF photonics; to our knowledge comparable levels are reported only in [27] and [35]. Furthermore, by incorporating our new nonlinear optical spectral smoothing approach into the tunable comb-based RF filter discussed earlier, in the future it should be possible to achieve high dynamic range lineshapes together with extremely rapid tunability, a combination very difficult to achieve using other approaches.

**Discussion**

An important point is that in our approach, filter lineshape is in principle independent of filter tuning; therefore, once sufficiently selective filter shapes are achieved, tuning should be possible with little additional control complexity. Rapid control of the RF bandwidth should also be possible. These operations could support a novel spectral zoom functionality, in which the RF spectrum is first scanned with a relatively coarse filter, which then zooms in to selected regions of interesting RF activity. Similar advantages should also extend to other types of filter response functions, including pulse compression filters. This is in marked contrast to conventional realizations of highly selective RF filters, which involve the coupled response of multiple resonances. Because of this coupling, tuning while maintaining highly selective lineshapes is difficult, and control is challenging.

Our approach is based fundamentally on the concept of an RF photonic link, modified to realize filtering action – specifically demonstrating a route to simultaneous tuning, bandwidth control, and high MSSR and stopband attenuation. Due to inefficient RF→optical→RF transduction, achieving good RF link performance has traditionally been challenging. Our filters are subject to the same challenges and currently exhibit approximately 40 dB RF insertion loss at 0.5 mA photodetector current, which is not good enough for practical applications. However, using special low $V_\pi$ modulators and high current handling detectors, dramatically improved RF photonic links have been realized in the last several years[43,44], with RF gain >10 dB and noise figure of <6 dB demonstrated at frequencies beyond 10 GHz[43]. Analysis shows that the RF gain and noise figure of a short optical link based on modulation of a frequency comb obey the same expressions as links based on a single laser[45]. Furthermore, for fiber remoting applications involving long RF photonic transmission links, the use of comb sources has recently been shown to suppress Brillouin scattering, enabling higher optical power and improved RF link performance[46]. Therefore, with appropriate investment in state-of-the-art components, in the future it should be possible to realize comb-based RF photonic filters both with the novel tuning, reconfiguration, and filter selectivity properties introduced here and with significantly improved link performance.

In addition to direct electro-optic comb generation, our comb-based RF photonic filtering scheme may also be implemented with other comb sources synchronizable to an RF drive. These include recirculating frequency shifters as in [35], scaled to higher comb spacing by using a single sideband electro-optic modulator in an amplifying loop[47], as well as integrated Fourier domain mode-locked lasers proposed in [48]. Such sources may offer increased comb power that facilitate improvements in RF gain. Finally, sampling considerations limit the frequency span of our RF photonic filter that is free from Nyquist spurs to half of our 10-GHz comb repetition frequency[14] (see Methods). By employing higher frequency optical modulators already developed for optical communications, we can anticipate increasing comb spacings to at least 40 GHz, boosting the spur-free range of comb-based RF photonic filters by at least a factor of four.

In summary, we have demonstrated several new principles for RF photonic filters based on electro-optically generated frequency combs. A novel nonlinear optics approach enhances both the bandwidth and shape of combs, leading to >60 dB MSSR and >70 dB RF stopband attenuation in a fixed filter geometry. Switching of optical delay under electronic control leads to extremely rapid (~40 ns) tuning of the filter response, while proof-of-concept experiments point the way to potentially rapid bandwidth reconfigurability.

## METHODS

### FILTER COMPONENTS

Our RF photonic filters were constructed using a single sideband modulator, a 5.1 km length of dispersion compensating fiber with specified dispersion of -1259.54 ps/nm, and a 20-GHz bandwidth photodetector. For the bandwidth reconfiguration experiments of Fig 4, additional fiber was used which increased the overall dispersion by 31.7%, which reduced the FSR of the filter from 10.4 GHz to 7.9 GHz. RF connections were made with conventional coax cables with specified 18 GHz bandwidth. Single sideband modulation is achieved using a dual drive lithium niobate Mach-Zehnder modulator, which may be viewed as an interferometer with independent phase modulators in its two arms[49]. By driving the phase modulators with 90° RF phase shift and biasing the interferometer at 90° optical phase shift, one of the modulation sidebands can be canceled, leaving just a single sideband on each comb line. Our dual drive modulator has a half-wave voltage of 4.5 V at 1 kHz and a 3-dB bandwidth of 12.2 GHz. The phase shift between the RF drive signals is provided by a 90° hybrid splitter with a bandwidth of 1–12.4 GHz and a maximum phase imbalance of 7°. For the tunable filter experiments, we also suppress the carriers by using a periodic optical filter with nulls positioned at the comb frequencies. This is realized using a micro-optic interferometer with ~100 ps delay imbalance, implemented as a commercial differential phase shift keying (DPSK) demodulator.

**MEASUREMENT TECHNIQUES**

Comb spectra were measured using an optical spectrum analyzer (OSA) with 0.01 nm resolution. RF filter frequency response measurements were performed using a vector network analyzer (VNA), usually set for 10 Hz resolution bandwidth. Before plotting, frequency responses were normalized by subtracting out the response measured in a fixed filter configuration but without the dispersive fiber. This takes out RF frequency dependences due to cable loss and high frequency component roll-off, thereby focusing on the RF response intrinsically related to our photonic processing scheme[14].

Because of the periodic nature of discrete-time filters, unwanted higher-order mixing terms may also be generated by this filter. For instance, an optical sideband may mix with a neighboring comb line to produce a signal that is located in a higher frequency band. As discussed in detail elsewhere[14], a variety of tones at frequencies $m\Delta f$ and $m\Delta f \pm f_{RF}$ may be present at the output, where $f_{RF}$ is the frequency of a single RF input tone, $\Delta f$ is the comb spacing, and $m$ is an integer. In order to avoid aliasing related to such effects, sampling theory restricts operation of digital filters to an upper frequency of one half of the sampling frequency. In the case of an optical comb-based RF photonic filter, this sampling rate corresponds to the comb repetition rate $\Delta f$. Note that in a network analyzer measurement, which is based on coherent mixing with a reference tone, only a signal with frequency equal to that transmitted from the network analyzer is acquired; other output tones are not detected. Nevertheless, in order to avoid spurious output tones, the frequency span should be restricted to fall within a single "Nyquist zone," where the various Nyquist zones run from $p\Delta f/2$ through $(p+1)\Delta f/2$, where $p$ is a nonnegative integer. For figures in which the filter response is plotted over a range exceeding $\Delta f/2$, the boundaries between Nyquist zones are depicted as dashed vertical lines, similar to [14]. The free spectral range (FSR) of our filters may be adjusted though choice of the fiber dispersion. For the data of Figs. 1 and 6, due the length of available fiber, the filter FSRs fall close to 10 GHz, which is the boundary between second and third Nyquist zones. For the data of Fig. 4, for which additional fiber was available, the FSR is approximately 7.9 GHz, which places a filter peak near the center of the second Nyquist zone.

Experiments demonstrating rapid filter reconfiguration utilized a 24 Gsample/s RF arbitrary waveform generator to generate the multiple frequency RF input signal and a 50 Gsample/s real-time oscilloscope to digitize the time domain filter output. In time domain experiments, sampling effects are directly observable in the recorded traces when the measurement band exceeds one Nyquist zone. For the rapid tuning data reported in Fig. 3, an analog lowpass filter with 4-GHz bandwidth is employed to reject signals from other Nyquist zones. The time-frequency distribution of the filtered signal, Fig. 3(d), was highlighted by computing a spectrogram of the oscilloscope data using a sliding FFT window of 20.4 ns duration, with contour lines drawn at 5-dB intervals down to 20 dB.

**DUAL COMB FILTER TUNING EXPERIMENTS**

Our filter tuning experiments utilize a pair of nominally identical comb generators designed for approximately Gaussian spectra. From Eq. (2), the filter response is sensitive to the difference in the frequency dependent optical phase of the combs. Minimum filter bandwidth is obtained if the combs are identical so that their spectral phases cancel. Although simulations indicate that the filter shape is robust against minor differences in the drive conditions of the two comb generators, nevertheless, we attempt to generate two combs that are as close to identical as possible. The modulation indices on the phase modulators are matched by monitoring and comparing the individual optical spectra, and the indices of individual intensity modulators are set by monitoring their outputs on an RF spectrum analyzer as the bias conditions are changed. Next the RF phase (delay) between modulators is set by observing the comb with both RF and optical spectrum analyzers. By watching for certain key signatures (e.g., a minimized comb line or harmonic), it is possible to very accurately set the proper bias and delay conditions and make the two combs very similar in terms of amplitudes and phases.

In the filter tuning experiments, propagation in the dispersive fiber leads to a constant delay, or latency, between the control signal sent to the filter and the filter response. This latency is measured by connecting a continuous 4 GHz RF tone to the input of the filter (i.e., to the single sideband modulator) and a low frequency (10 kHz) square wave (~50 ps rise time, 1 V amplitude) to the control port of the filter (i.e., to the RF phase shifter). This scheme switches the filter passband alternately to coincide with or to miss the 4 GHz input, leading to a strong contrast in the output which is observed with a real-time oscilloscope synchronized to the low frequency square wave. This measurement yields a latency of 26.0 μs.

# COMB GENERATION BY CASCADED FOUR WAVE MIXING

In the comb source of Fig. 5(a), the two CW lasers are spaced 10 nm apart. We operate the RF oscillator at 10 GHz, for which the $V_\pi$ is ~9 V for the IM's and ~3 V for the PM's. IM1 is biased at quadrature with the RF amplitude of 0.5 $V_\pi$ zero to peak. IM2 is biased at the maximum transmission point with the RF amplitude of $V_\pi$. The transmission function is $a_1=0.5[1+\exp(-j0.5\pi)\exp(j0.5\pi\cos(\omega_{RF}t))]$ for IM1, and $a_2=0.5[1+\exp(j\pi\cos(\omega_{RF}t))]$ for IM2. The phase modulation $\phi(t)$ is from the single phase modulator, driven at its maximum RF input power (30 dBm) to maximize the modulation index seen by the gated pulses after IM2. The SMF used to delay one frequency with respect to the other by half a modulation period is ~300m. This is followed by a 1.5-W optical amplifier and a 100 m length of highly nonlinear fiber, with near zero dispersion and low dispersion slope, to create a cascade of four-wave mixing (FWM) terms. By use of a bandpass filter, we select the FWM term centered at 1562 nm. More details on the broadband comb generation process using cascaded FWM between two phase modulated CW lasers can be found in [50].

**Acknowledgements**

This project was supported in part by the Naval Postgraduate School under grant N00244-09-1-0068 under the National Security Science and Engineering Faculty Fellowship program. Any opinion, findings, and conclusions or recommendations expressed in this publication are those of the authors and do not necessarily reflect the views of the sponsors.

We would like to acknowledge Dr. Victor Torres-Company, Dr, Jason McKinney, and Prof. Dimitrios Peroulis for helpful discussions.

**Competing financial interests**

The authors declare that they have no competing financial interests.

**Materials & Correspondence**

Correspondence and requests for materials should be addressed to A.M. Weiner.

**Statement of author contributions**

**Figure 1 – Representative operation of an optical frequency comb based multi-tap RF photonic filter**

**a,** Schematic of a multi-tap microwave photonic filter using an optical frequency comb, SSB – Single sideband modulation, DCF – Dispersion compensating fiber, AMP – fiber amplifier, PD – Photodetector, **b,** Experimental scheme and **c,** pictorial representation of direct quasi-Gaussian frequency comb generation, CW – continuous-wave laser, IM – Intensity modulator, PM – Phase modulator, PS – Microwave phase shifter, **d,** Optical spectrum of generated frequency comb**,** and **e,** Measured (blue) and simulated (red) RF filter response using this comb. The vertical dashed lines depict individual Nyquist zones, explained under Methods.

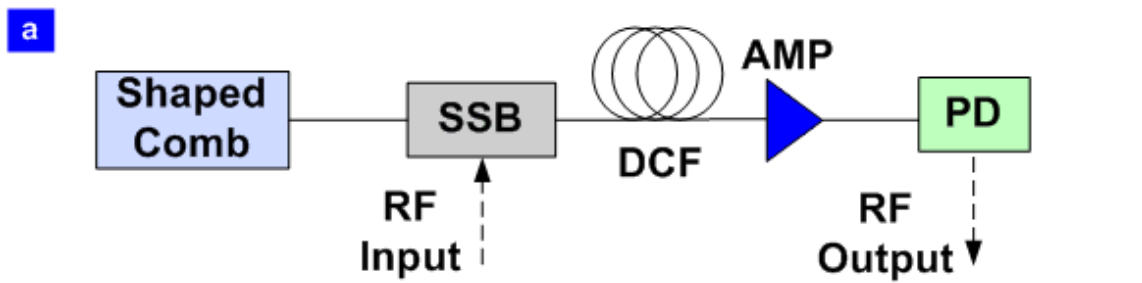

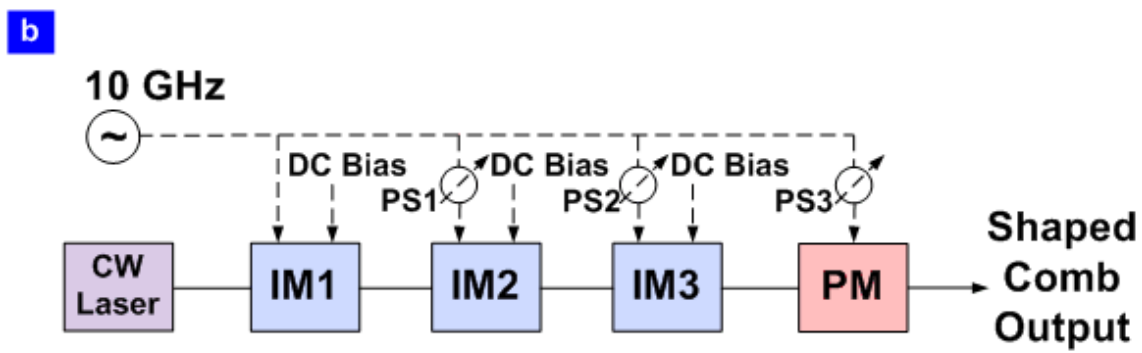

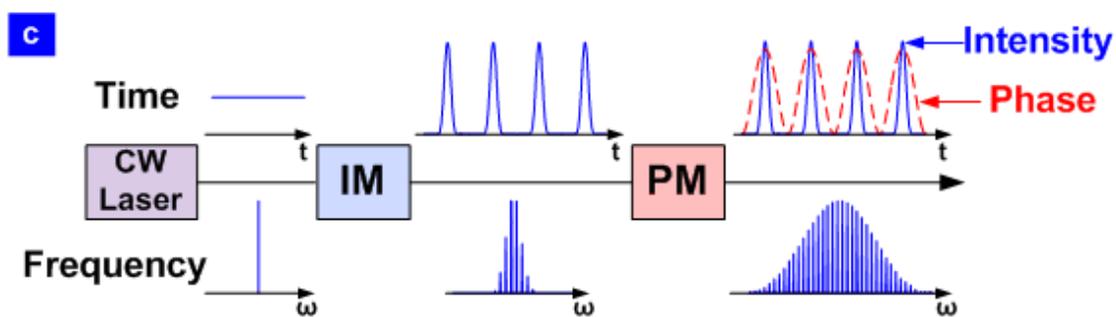

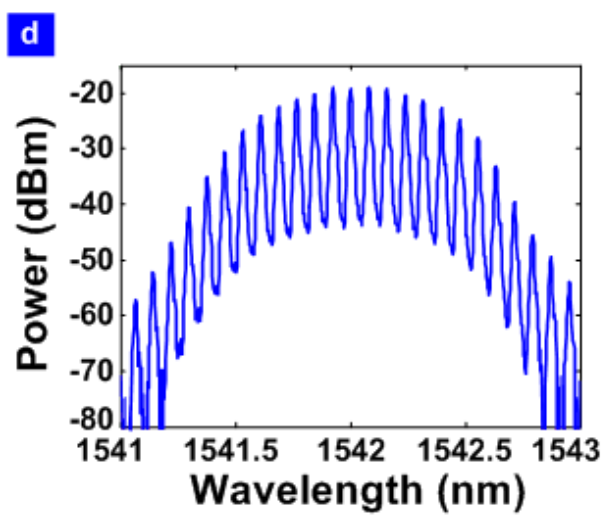 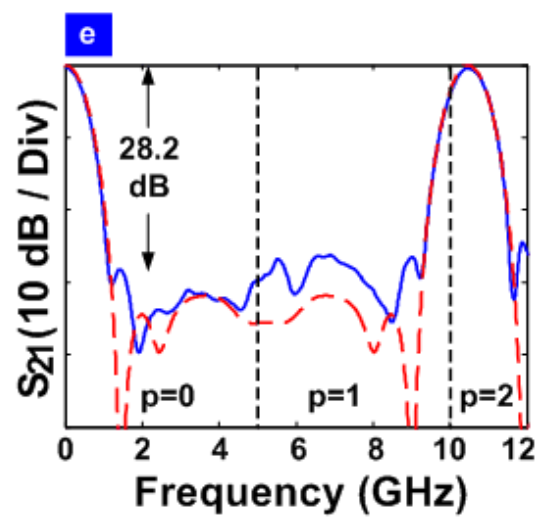

**Figure 2 – Achieving rapid electronic frequency tunability of the RF filter using a dual-comb approach.**
**a**, Scheme depicting the use of optical delay between the combs to achieve filter tunability, SSB-SC – Single sideband modulation with suppressed carrier, **b,** Method to achieve rapid tunability using a dual comb scheme, **c,** Experimental results showing the programmable rapid delay tuning between the two combs by applying a rapid step to the RF phase shifter, **d,** Tuning the center frequency of the RF filter by changing the voltage to the phase shifter (quasi-static case)

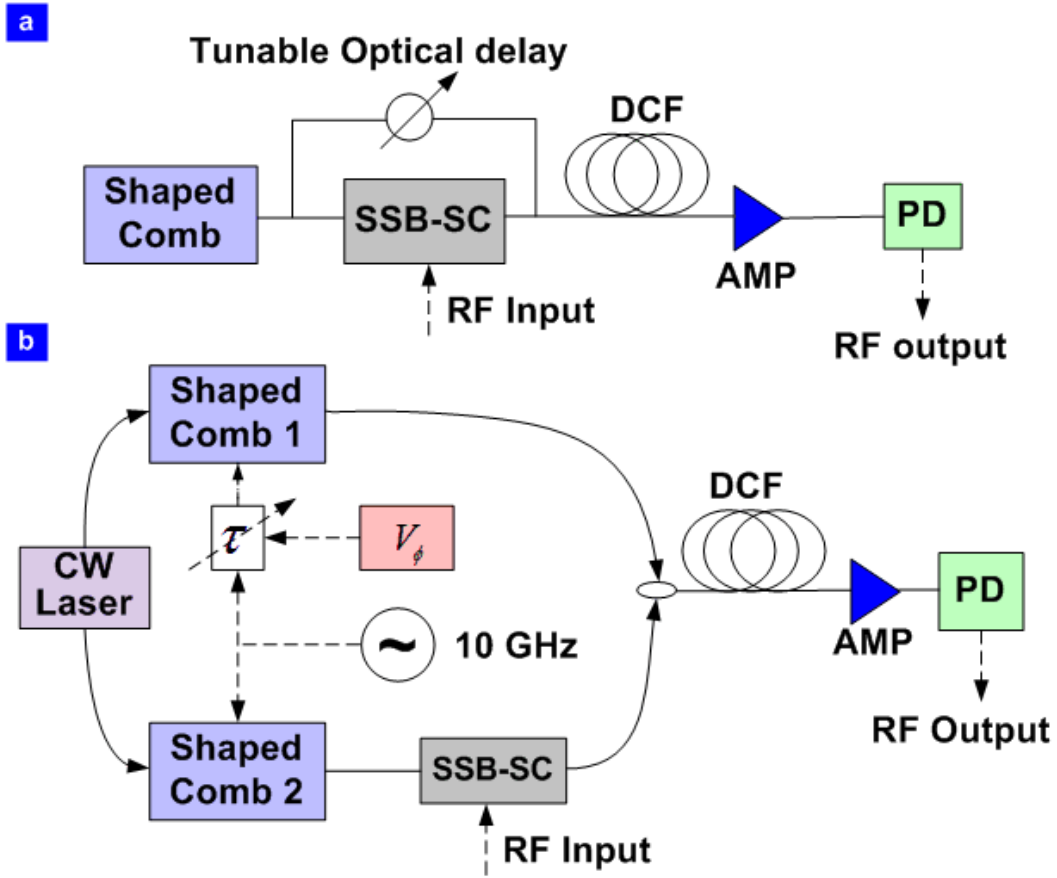
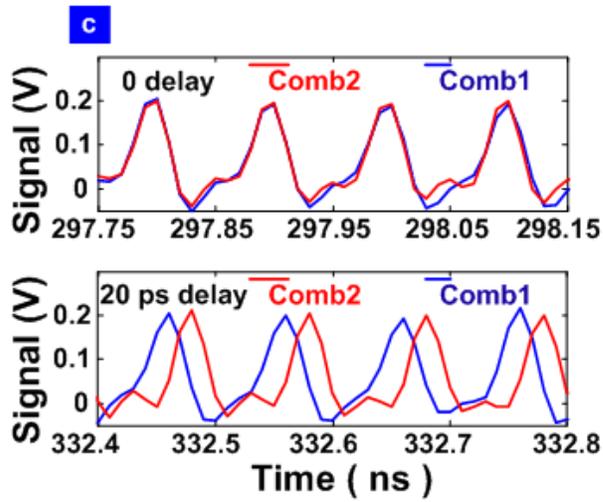
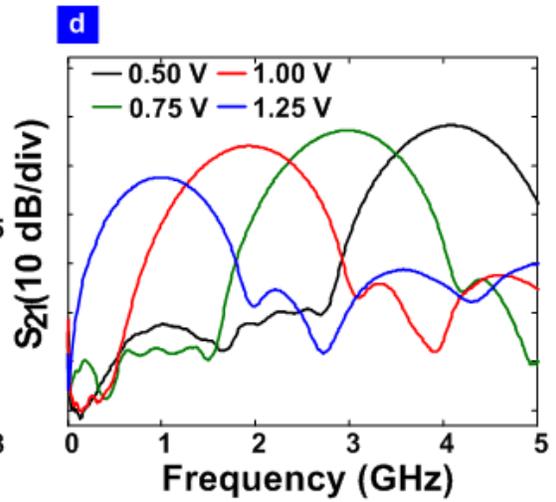

**Figure 3 – Rapid frequency tunability of the RF filter.**

**a,** Measured oscilloscope signal demonstrating rapid switching between different frequency bands. Also shown is the control signal input to the phase shifter. **b,** Close up of a first waveform part of the signal - a 3.7 GHz RF tone **c,** Close up of a second waveform section – a 1.8 GHz RF tone undergoing binary phase shift keying at 600 Mb/s. **d,** Spectrogram representation of the filtered signal, highlighting rapid switching between different frequency bands.

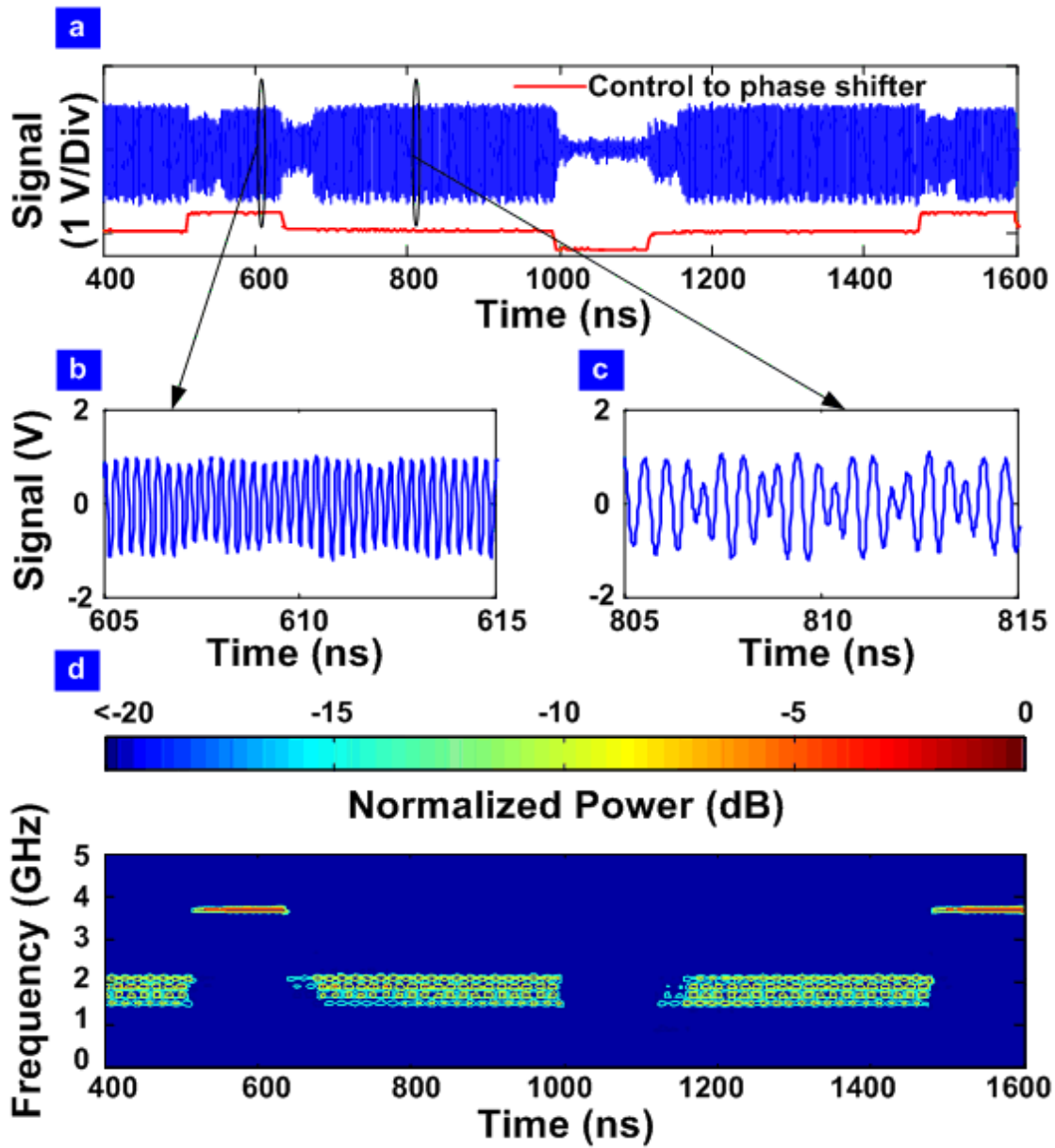

**Figure 4 – Demonstration of bandwidth tunability of the RF filter**

**a**, Scheme for modifying the bandwidth of the generated optical comb, VA – variable RF attenuator, **b**, RF filter passbands for the different combs for different drive voltages, with red dashed curve corresponding to approximately a factor of two reduced drive level compared to blue solid line indicating bandwidth tuning with a factor of two. BW - Bandwidth.

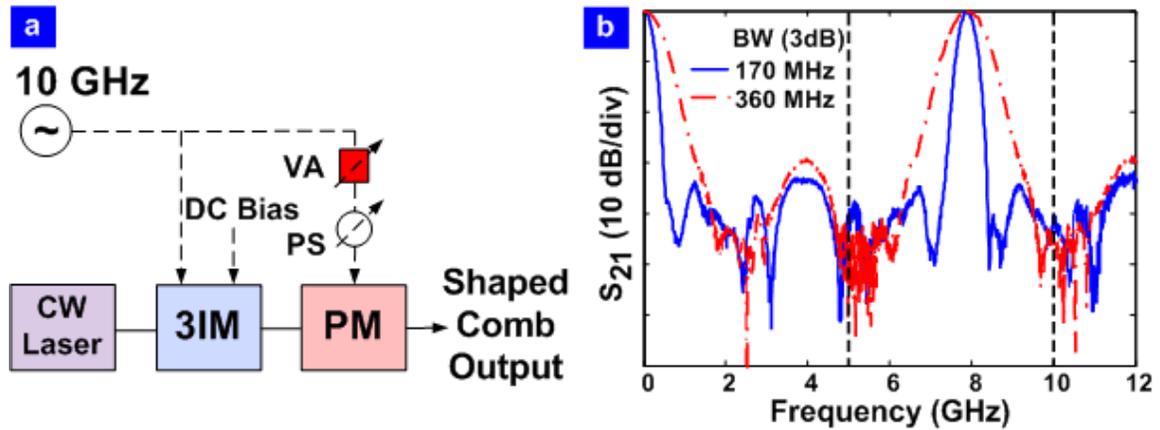

**Figure 5 – Direct, high accuracy generation of Gaussian apodized broadband combs**
**a**, Experimental scheme to directly generate broadband, apodized combs using cascaded four-wave mixing in highly nonlinear fiber. The electro-optic modulator portion of the comb generator is similar to Fig. 1(b), but simplified by reducing the number of intensity modulators. SMF – single mode fiber, HNLF – Highly nonlinear fiber. **b** Depiction of the cascaded four-wave mixing process. Successive terms scale in bandwidth and correspond increasingly accurately to a smooth Gaussian apodization.

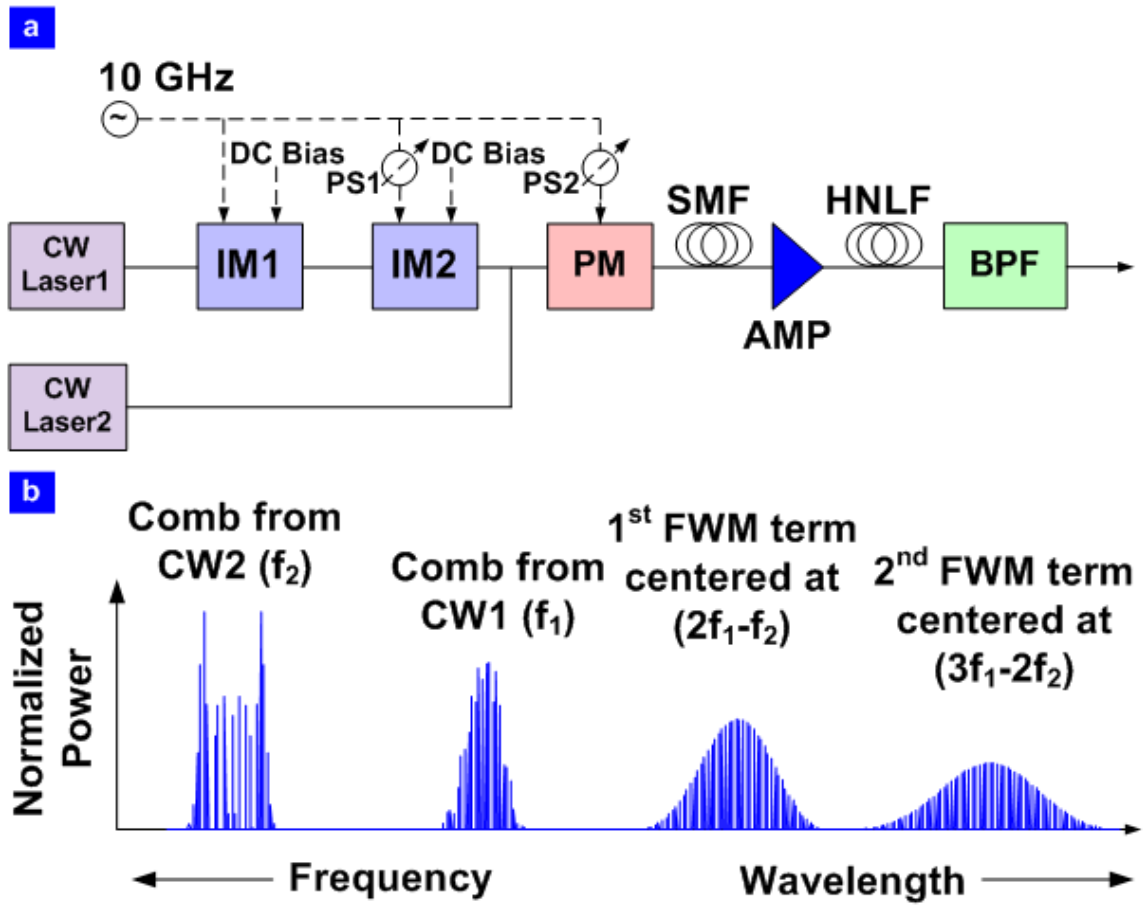

**Figure 6 – Achieving high stop-band attenuation RF filters**

**a**, Spectrum from laser CW1 measured directly after the modulators, **b,** Measured spectrum of the 2$^{nd}$ order cascaded FWM term used in our experiment. **c,** Measured RF filter transfer function using the 2$^{nd}$ order FWM term, Fig. 5(b), demonstrating extremely high main lobe to sidelobe suppression ratio (61 dB) and stop-band attenuation (>70 dB). Also shown is the measured transfer function using only the comb after the modulators, Fig. 5(a).

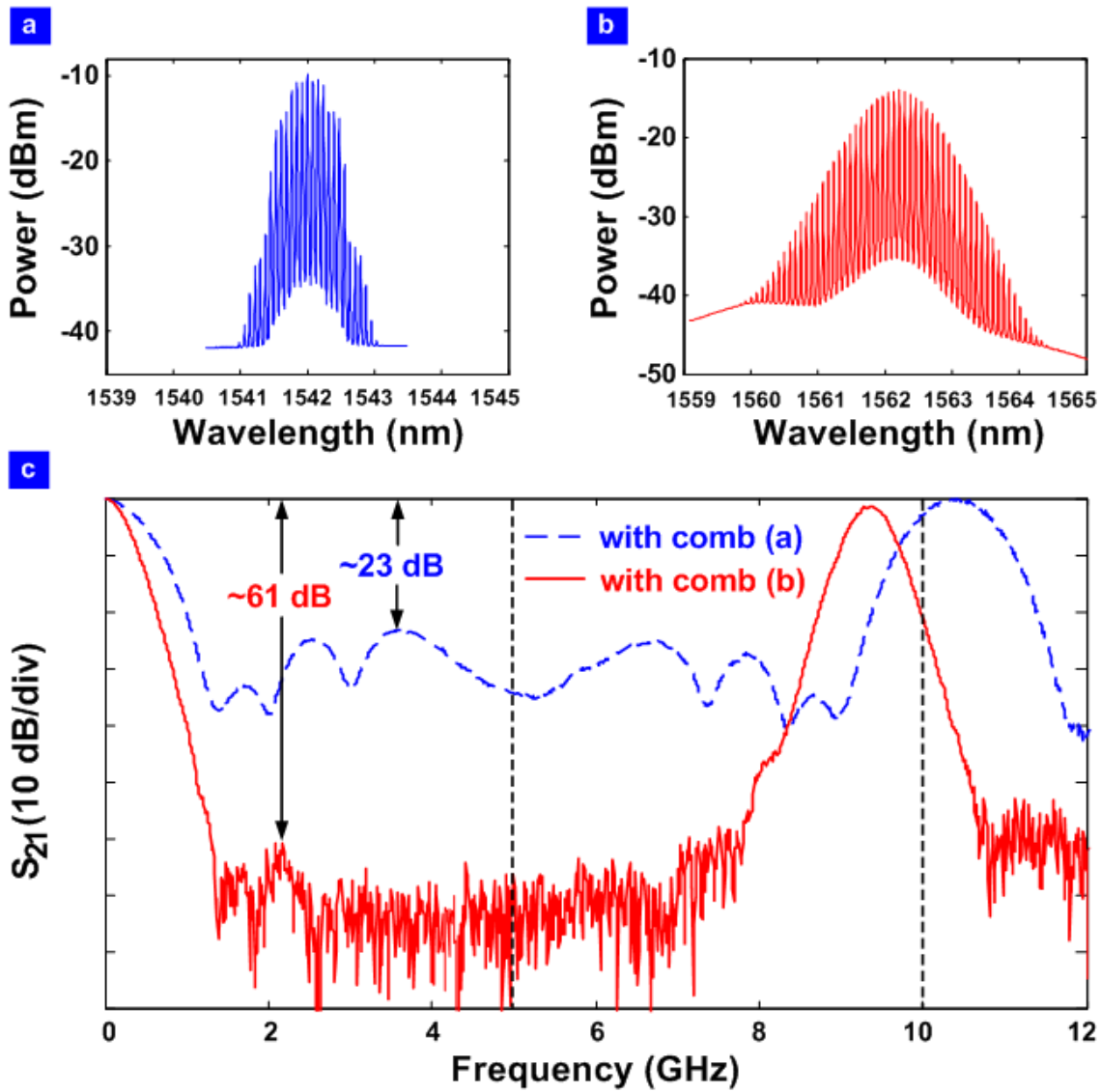